\documentclass[aps,twocolumn,superscriptaddress,floatfix,longbibliography]{revtex4-2}
\usepackage{amsmath,amssymb,amsthm}
\usepackage{physics}
\usepackage{amsfonts}
\usepackage{mathrsfs}
\usepackage{graphicx}
\usepackage{tabularx}
\usepackage{enumerate}
\usepackage{dcolumn}
\usepackage{bm}
\usepackage{xcolor}
\usepackage[normalem]{ulem}
\usepackage[colorlinks,linkcolor=blue,citecolor=blue,urlcolor=blue]{hyperref}

\begin{document}

\title{Dissipation-tunable extended and localized steady states in a non-disordered lattice}

\author{Ming-Jie Tao}
\email{taomingjie1020@sina.com}
\affiliation{College of Mathematics and Physics, Chengdu University of Technology, Chengdu 610059, China}

\author{Yi-Ting Wang}
\affiliation{College of Mathematics and Physics, Chengdu University of Technology, Chengdu 610059, China}

\author{Jing Li}
\affiliation{College of Mathematics and Physics, Chengdu University of Technology, Chengdu 610059, China}


\author{Hongsheng Hou}
\affiliation{School of Physics, Hangzhou Normal University, Hangzhou, Zhejiang 311121, China}


\author{Xiang-Ping Jiang}
\email{2015iopjxp@gmail.com}
\affiliation{School of Physics, Hangzhou Normal University, Hangzhou, Zhejiang 311121, China}

\author{Lei Pan}
\email{panlei@nankai.edu.cn}
\affiliation{School of Physics, Nankai University, Tianjin 300071, China}

\date{\today}

\begin{abstract}

Dissipation is usually regarded as a source of decoherence that suppresses quantum interference and localization. Here we show that suitably engineered dissipation can instead be used to select localized or extended states in a strictly non-disordered one-dimensional lattice. The underlying clean lattice has spatially inhomogeneous hopping and supports both extended bulk states and localized boundary states, including an algebraically localized bound state in the continuum. We introduce a nonlocal bond jump operator with a tunable relative phase and show that this phase selectively favors eigenstates with different spatial phase correlations. As a result, the long-time density matrix can be steered toward sectors dominated by localized or extended Hamiltonian eigenstates without changing any Hamiltonian parameter. The microscopic origin of the selection is quantified by the fraction of site pairs separated by a distance $l$ that are phase matched with the dissipative channel. We further characterize the dissipative quench through the quantum fidelity and show that the selected character of the steady state can persist after the dissipation is removed. Our results establish phase-selective bond dissipation as a route to controllable state preparation and transport manipulation in non-disordered lattices.

\end{abstract}

\maketitle

\section{Introduction}

The phenomenon of Anderson localization, first elucidated by P.~W. Anderson in 1958~\cite{anderson1958absence}, fundamentally revolutionized our understanding of wave propagation in disordered media. It established that random potential fluctuations can exponentially localize electron waves, completely suppressing spatial diffusion and inducing a metal-insulator transition~\cite{thouless1974electrons,abrahams1979scaling,lee1985disordered,kramer1993localization,evers2008anderson,hetenyi2021scaling}. Over the subsequent decades, this seminal framework has inspired extensive investigations across diverse physical platforms, ranging from photonic architectures and acoustic materials to ultracold atomic gases. However, it was shown that localization can also occur in non-disordered systems~\cite{chalker2010anderson,molina2012surface,corrielli2013observation,zeng2024transition}, challenging the traditional paradigm that randomness is essential for wave localization. These findings include localization in clean lattices with tailored nonlinearities or through engineered synthetic dimensions, offering new pathways to control wave propagation without relying on disorder~\cite{leykam2013flat,bodyfelt2014flatbands,leykam2018artificial,jin2019flat}.

Concurrently, the field of open quantum systems has experienced a remarkable resurgence, propelled by experimental advancements in the precise engineering of dissipation and system-environment dynamics~\cite{prosen2008quantum,mebrahtu2012quantum,longhi2019topological,hamazaki2019non,xu2020topological,liuT2020non,shastri2020dissipation,nie2021dissipative,yamamoto2021collective,zeng2020topological1,zeng2020topological2,weidemann2022topological,wu2021non,li2023non,zhu2023topological,liuT2020non,gandhi2023topological,liu2023ergodicity,kawabata2023entanglement,li2024emergent,yu2024non,zhou2024entanglement1,zhou2024entanglement2,jing2024biorthogonal,yang2026noise,yang2026noise1}. In this context, the interplay between environmental dissipation and wave transport has become a focal point of research~\cite{gurvitz2000delocalization,yamilov2014position,huse2015localized,balasubrahmaniyam2020necklace,weidemann2021coexistence,purkayastha2017nonequilibrium,vershinina2017control,yusipov2018quantum,vakulchyk2018signatures,balachandran2019energy,chiaracane2020quasiperiodic,lacerda2021dephasing,chiaracane2022dephasing,dwiputra2021environment,longhi2023anderson,jiang2024exact1,peng2024dissipation,xu2026dissipative,wei2026quantum,song2026quantum,wei2026symmetry,xu2026strong}. The traditional consensus posits that dissipation is inherently detrimental to quantum interference; environment-induced dephasing is generally expected to erode the precise phase relationships required for localized states, inevitably driving the system toward an ergodic, delocalized steady state. Yet, emerging theoretical studies suggest a much more nuanced role for dissipation. For instance, it has been shown that tailored dissipation can drive Anderson localization into a robust stationary state without destroying it~\cite{yusipov2017localization,liu2024dissipation,longhi2024dephasing}, and can even induce dephasing-driven mobility edges (MEs) in quasicrystals. Furthermore, in one-dimensional quasiperiodic systems featuring MEs, specific dissipative couplings can mediate deterministic transitions between extended and localized states~\cite{yang2025dissipation,yang2025dissipation1,xu2026dissipation,jiang2026dissipation}. Despite these advances, a critical gap remains: the capacity of dissipation to actively induce and control extended-to-localized transitions in strictly non-disordered systems remains largely unexplored.

In this work, we address this gap by exploring a tunable transition between extended and localized steady states in a non-disordered lattice, driven entirely by engineered bond dissipation. By constructing phase-selective dissipative jump operators, we demonstrate precise, deterministic control over the asymptotic states and transport characteristics of the open system. We systematically investigate the steady-state density matrix distributions in both real-space and eigenstate representations, and track the corresponding dissipative dynamical evolution. Our findings reveal that, irrespective of the initial preparation, the system can be robustly steered into a stationary regime dominated by either extended or localized modes. Crucially, this transition is governed by experimentally accessible parameters of the dissipative coupling: the relative phase $\alpha$ and the inter-site distance $l$. This work establishes non-local bond dissipation as a versatile and powerful tool for tuning localization transitions, offering new theoretical perspectives on quantum state manipulation in open, disorder-free architectures.

The remainder of this paper is organized as follows: In Sec.~\ref{sec: models}, we define the non-disordered tight-binding model featuring inhomogeneous hopping rates that sustains bound states in the continuum. In Sec.~\ref{sec: Lindblad}, we describe the open-system dynamics using the Lindblad master equation equipped with phase-dependent bond dissipator. The numerical results concerning the dissipation-tunable extended and localized steady states are analyzed and discussed in Sec.~\ref{sec: results}. Finally, a comprehensive summary of our findings is provided in Sec.~\ref{sec: conclusion}.

\section{The model Hamiltonian}\label{sec: models}

\begin{figure}[t!]
	\centering
	\includegraphics[width=0.48\textwidth]{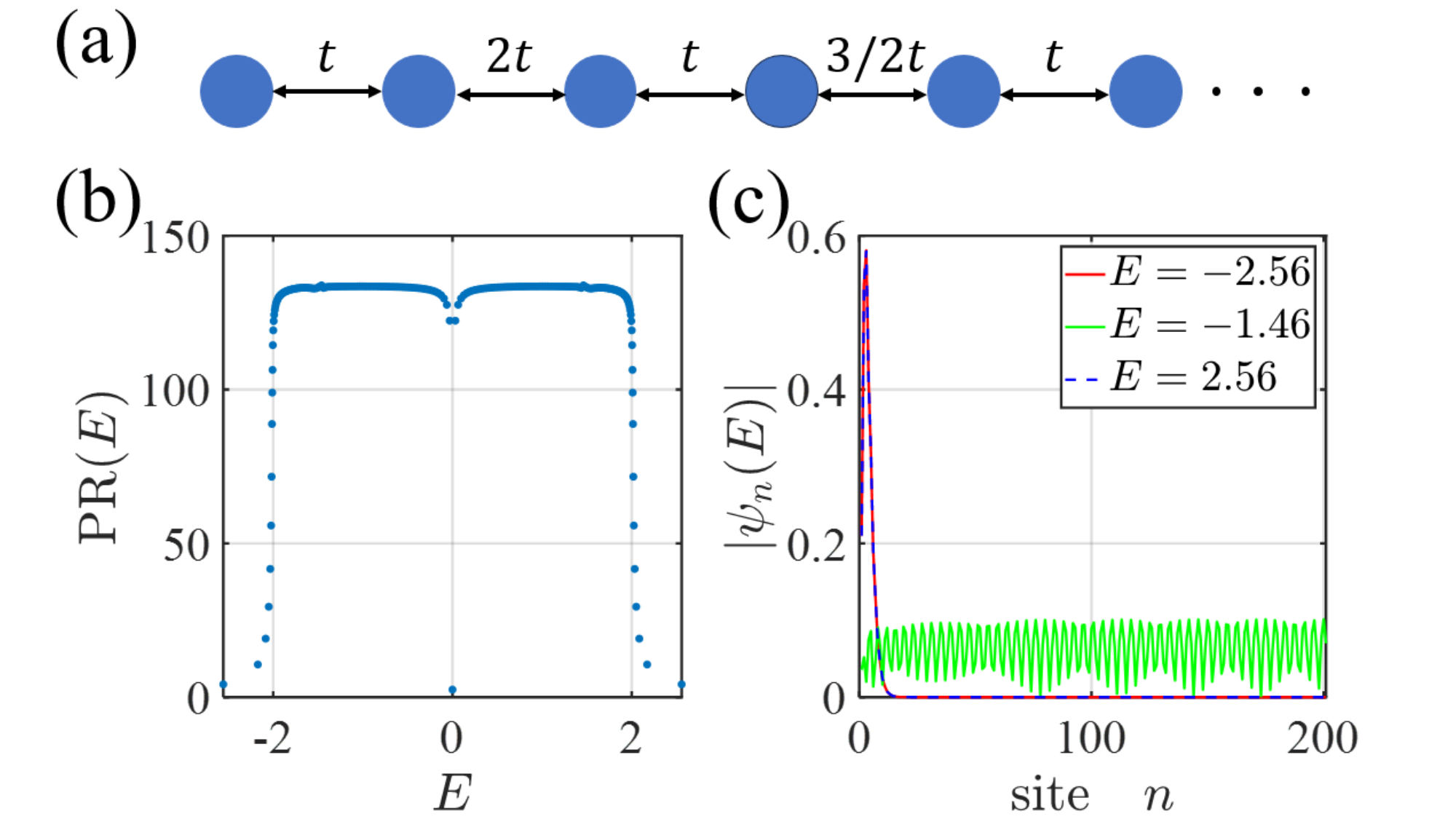}
        \caption{(a) Schematic of the inhomogeneously-hopped lattice with $N=2$ hopping modulation, which supports $M=1$ surface bound state in the continuum (BIC) at the parameter point $\beta=1$. (b) Participation ratio ${\rm PR}(E)$ of the single-particle eigenstates versus their energy $E$, computed numerically for a finite lattice of $L=201$ sites; the horizontal spread of the data simultaneously displays the energy spectrum. (c) Real-space wavefunction amplitudes $|\psi_n(E)|$ for representative eigenstates: an extended bulk state at $E=-1.46$, delocalized across the lattice, and localized boundary modes at $E=\pm 2.56$, confined near the lattice edges.}
	\label{fig1}
\end{figure}

We consider a strictly disorder-free one-dimensional (1D) tight-binding lattice with spatially modulated nearest-neighbor hopping amplitudes. The corresponding non-disordered Hamiltonian is
\begin{equation}\label{equation1}
H=\sum_{j=1}^{L-1}\left(t_j c_{j+1}^{\dagger}c_j+\mathrm{H.c.}\right),
\end{equation}
where $c_j^\dagger$ ($c_j$) denotes the creation (annihilation) operator at site $j$, and $t_j>0$ is the real hopping amplitude between sites $j$ and $j+1$. Rather than introducing disorder through random onsite potentials or quasiperiodic modulations, we consider a deterministic spatial modulation of the hopping amplitudes that is localized near the surface and approaches the uniform bulk value $t$ away from the boundary. Specifically, we define
\begin{equation}\label{equation2}
t_j= \left\{ 
\begin{array}{ll}
t, &  j \neq \kappa N, \\
\left( \frac{\kappa+1}{\kappa} \right)^{\beta}t, & j=\kappa N \; \; (\kappa=1,2,3,...),
\end{array}
\right.
\end{equation}
where $N=M+1$ and $\beta$ is a real parameter controlling the algebraic modulation of the hopping amplitudes, as shown in Fig.~\ref{fig1}(a). In the thermodynamic limit, $t_j\rightarrow t$ as $j\rightarrow\infty$, such that the lattice remains asymptotically homogeneous in the bulk. The single-particle eigenvalue equation corresponding to Eq.~(\ref{equation1}) can be written as
\begin{equation}
\label{equation3}
E\psi_j=t_j\psi_{j+1}+t_{j-1}\psi_{j-1},
\end{equation}
where $\psi_n$ denotes the wave-function amplitude at site $j$. Since the lattice is asymptotically uniform, its continuous spectrum is given by the bulk tight-binding band $-2t<E<2t$. The spatial modulation of the hopping amplitudes can, however, generate localized states whose energies either lie within or outside this continuum. In particular, the model supports $M$ surface bound states embedded in the continuous spectrum. Their wave functions can be expressed as
$\psi_j^{(\sigma)}=A_j\sin(jk_\sigma)$ with $E_\sigma=-2t\cos k_\sigma,\qquad k_\sigma=\frac{\pi\sigma}{N}, \qquad \sigma=1,2,\ldots,M.$ Here, the amplitude factor is piecewise constant, $A_j=\mathcal{N}\kappa^{-\beta}, \qquad (\kappa-1)N<j\leq \kappa N,$ where $\mathcal{N}$ is determined by normalization. The resulting algebraic decrease of the envelope distinguishes these surface bound states from conventional exponentially localized boundary modes.

As an illustrative example, Fig.~\ref{fig1}(b) shows the numerically obtained spectrum of Eq.~(\ref{equation3}) for a finite lattice of $L=201$ sites with $N=2$ and $\beta=1$. In this case, $M=1$, and the model supports a single surface bound state in the continuum (BIC) at $E=0$. The corresponding wave-function envelope decays algebraically as $|\psi_n|\propto 1/(j+1)$. Thus, the state remains spatially localized despite its energy being embedded in the extended tight-binding band.

In addition to the surface BIC, the same hopping modulation produces localized states with energies outside the bulk continuum. These states are exponentially localized near the surface and therefore constitute bound states outside the continuum. For the parameters considered in Fig.~\ref{fig1}(b), the two outermost states occur at approximately $E\simeq\pm E_0$, with $E_0\simeq2.56t$, while additional bound states accumulate toward the band edges at $E=\pm2t$. The coexistence of a surface BIC with exponentially localized states outside the bulk band provides a useful setting for distinguishing different localization mechanisms in a deterministic, disorder-free lattice.

To quantify the spatial extent of the eigenstates, we use the participation ratio
\begin{equation}
\mathrm{PR}(E)=\frac{\left(\sum_j|\psi_j|^2\right)^2}{\sum_j|\psi_j|^4}.
\end{equation}
For a normalized state, $\mathrm{PR}$ measures the effective number of lattice sites occupied by the wave function. Accordingly, $\mathrm{PR}=O(1)$ for a strongly localized state, whereas $\mathrm{PR}=O(L)$ for an extended state. The spatial profiles of three representative eigenstates are shown in Fig.~\ref{fig1}(c), illustrating the coexistence of the algebraically localized surface BIC at $E=0$ and exponentially localized bound states outside the continuum. The BIC is also structurally robust against perturbations of the lattice parameters, as discussed below. This combination of localized surface states and extended bulk states in the absence of disorder provides a controlled platform for studying localization and phase-selective state transitions induced by dissipation.

\section{The Lindblad master equation and the bond dissipation}\label{sec: Lindblad}

To investigate the nonunitary dynamics and the emergence of robust steady states in our disorder-free platform, we formulate the problem as an open quantum system governed by the Lindblad master equation~\cite{lindblad1976generators}.
Within the Born--Markov approximation, the density matrix $\rho(t)$ evolves according to
\begin{equation}
\frac{d\rho(t)}{dt}=\mathcal{L}[\rho(t)]=-i[H,\rho(t)]+\mathcal{D}[\rho(t)],
\label{eq:master_eq}
\end{equation}
where $H$ is the tight-binding Hamiltonian of Eq.~\eqref{equation1}, which generates the coherent unitary dynamics, and $\mathcal{L}$ denotes the Liouvillian superoperator. The dissipator takes the standard Gorini--Kossakowski--Sudarshan--Lindblad form
\begin{equation}
\mathcal{D}[\rho(t)]=\Gamma\sum_{j}\left(
O_j\rho(t)O_j^\dagger-\frac{1}{2}\{O_j^\dagger O_j,\rho(t)\}
\right),
\label{eq:dissipator}
\end{equation}
with $O_j$ the jump operator of the $j$th dissipative channel and $\Gamma$ the overall dissipation rate. The Hamiltonian and dissipative terms in Eq.~\eqref{eq:master_eq} thus describe, respectively, the reversible coherent evolution and the irreversible coupling to a Markovian environment.

\begin{figure*}[t!]
	\centering
	\includegraphics[width=0.95\textwidth]{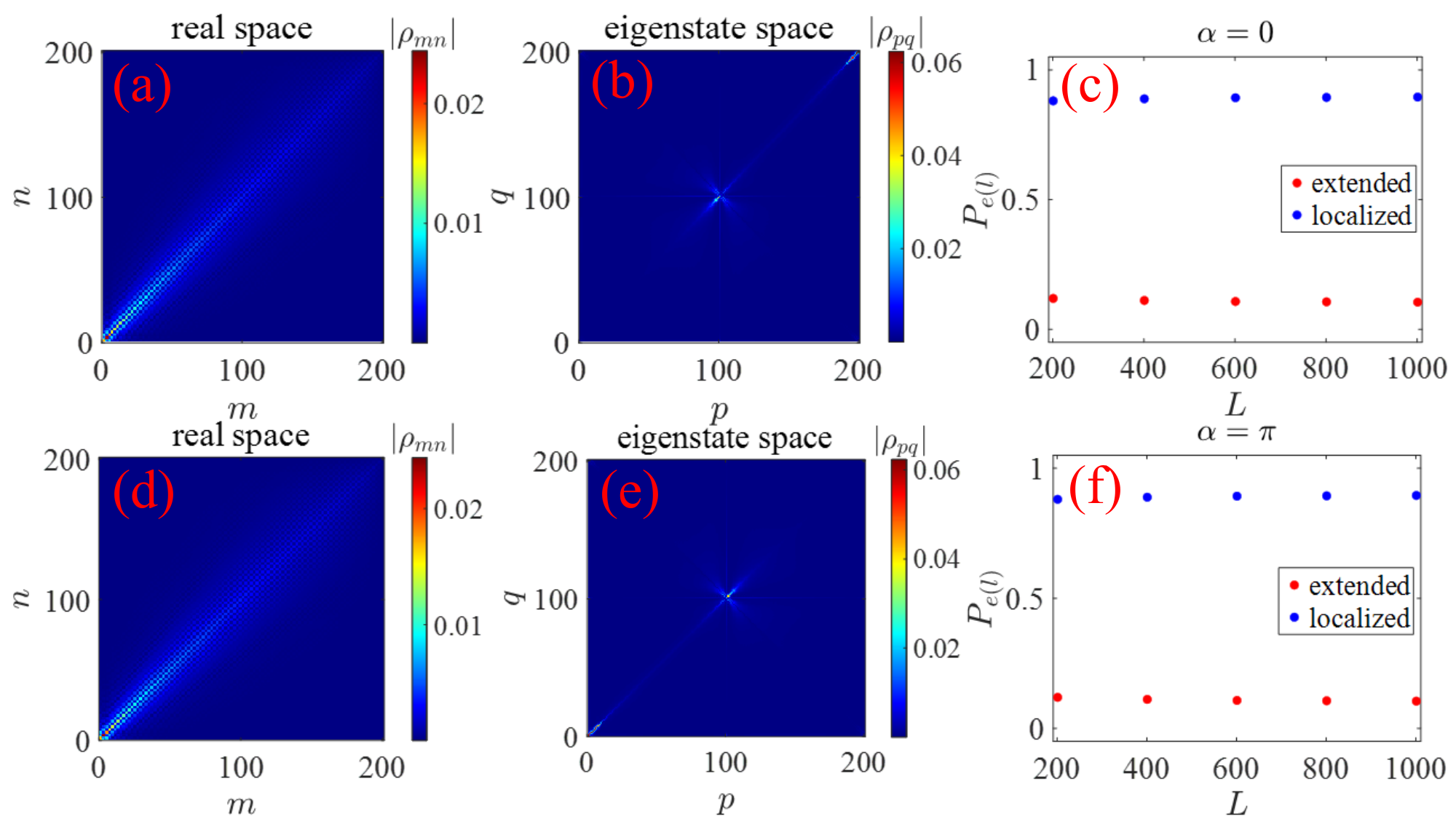}
        \caption{Localization properties of the stationary state for $\alpha=0$ and $\alpha=\pi$. (a),(d) Absolute values of the steady-state density-matrix elements $|\rho_{mn}|$ in real space for $\alpha=0$ and $\alpha=\pi$, respectively. (b),(e) Corresponding density-matrix distributions $|\rho_{pq}|$ in the eigenstate basis of the underlying Hamiltonian. (c),(f) Finite-size scaling of the eigenstate localization measure $P_e^{(l)}$ for system sizes $L=201$, $401$, $601$, $801$, and $1001$. The localized and extended states remain well separated over the entire size range for both values of $\alpha$, with no appreciable systematic drift with increasing $L$. The results demonstrate the robustness of the localized and extended sectors against the phase reversal of the dissipative channel. Other parameters are $l=1$, and $\Gamma=1$.}
	\label{fig2}
\end{figure*}

To determine the stationary solutions and the full relaxation spectrum, we vectorize the density matrix, mapping it onto a Liouville--Fock space via the Choi--Jamio{\l}kowski isomorphism~\cite{jamiolkowski1972linear,choi1975completely}.
Writing $\rho=\sum_{m,n}\rho_{mn}|m\rangle\langle n|$, we define
$|\rho\rangle\!\rangle=\sum_{m,n}\rho_{mn}|m\rangle\otimes|n\rangle
\in\mathcal{H}\otimes\mathcal{H}^{*}$. The master equation is thereby
cast as a linear evolution equation,
$\frac{d}{dt}|\rho(t)\rangle\!\rangle=\mathcal{L}|\rho(t)\rangle\!\rangle$,
in which the Liouvillian is represented by the non-Hermitian matrix
\begin{align}
\mathcal{L}=&-i\left(H\otimes\mathbb{I}-\mathbb{I}\otimes H^{T}\right)
\nonumber\\
&+\Gamma\sum_{j}\left(
O_j\otimes O_j^{*}
-\frac{1}{2}\,O_j^\dagger O_j\otimes\mathbb{I}
-\frac{1}{2}\,\mathbb{I}\otimes(O_j^\dagger O_j)^{T}
\right),
\label{eq:liouvillian_vec}
\end{align}
where $\mathbb{I}$ is the identity operator on $\mathcal{H}$. For a time-independent Liouvillian, the long-time behavior is controlled by the zero-eigenvalue subspace of $\mathcal{L}$; whenever this subspace is one-dimensional, the system relaxes to a unique steady state $\rho_{\rm ss}$ satisfying $\mathcal{L}|\rho_{\rm ss}\rangle\!\rangle=0$.

The spectral structure of $\mathcal{L}$ and the character of the resulting steady state are set by the microscopic form of the jump operators. In contrast to conventional local particle loss ($O_j=c_j$) or on-site dephasing ($O_j=c_j^\dagger c_j$), we engineer a nonlocal, phase-selective bond dissipation described by
\begin{equation}
O_j=\left(c_j^\dagger+e^{i\alpha}c_{j+l}^\dagger\right)
\left(c_j-e^{i\alpha}c_{j+l}\right).
\label{eq:jump_operator}
\end{equation}
This operator acts simultaneously on sites $j$ and $j+l$. It conserves the total particle number, $[\sum_j c_j^\dagger c_j,O_j]=0$, and therefore acts as a pure decoherence channel that enforces dissipative phase locking between the two sites. The phase $\alpha$ selects the target coherence pattern. For $\alpha=0$, the annihilation factor $(c_j-c_{j+l})$ vanishes on the antisymmetric single-particle mode $(c_j^\dagger-c_{j+l}^\dagger)|0\rangle$, while the creation factor $(c_j^\dagger+c_{j+l}^\dagger)$ projects onto the symmetric mode; the dissipator thus drives the two sites into in-phase
synchronization. For $\alpha=\pi$, the selection rule is reversed, and the system is driven into an out-of-phase (antisymmetric) configuration. As we demonstrate below, this continuous tunability of $\alpha$ enables the system to be steered between extended bulk states or localized boundary states. We note that nonlocal jump operators of this form are accessible in current engineered-reservoir platforms, having been proposed and investigated in optical Raman lattices~\cite{liu2024dissipation} and 1D Bose--Hubbard chains~\cite{marcos2012photon}.

\begin{figure*}[t!]
	\centering
	\includegraphics[width=0.95\textwidth]{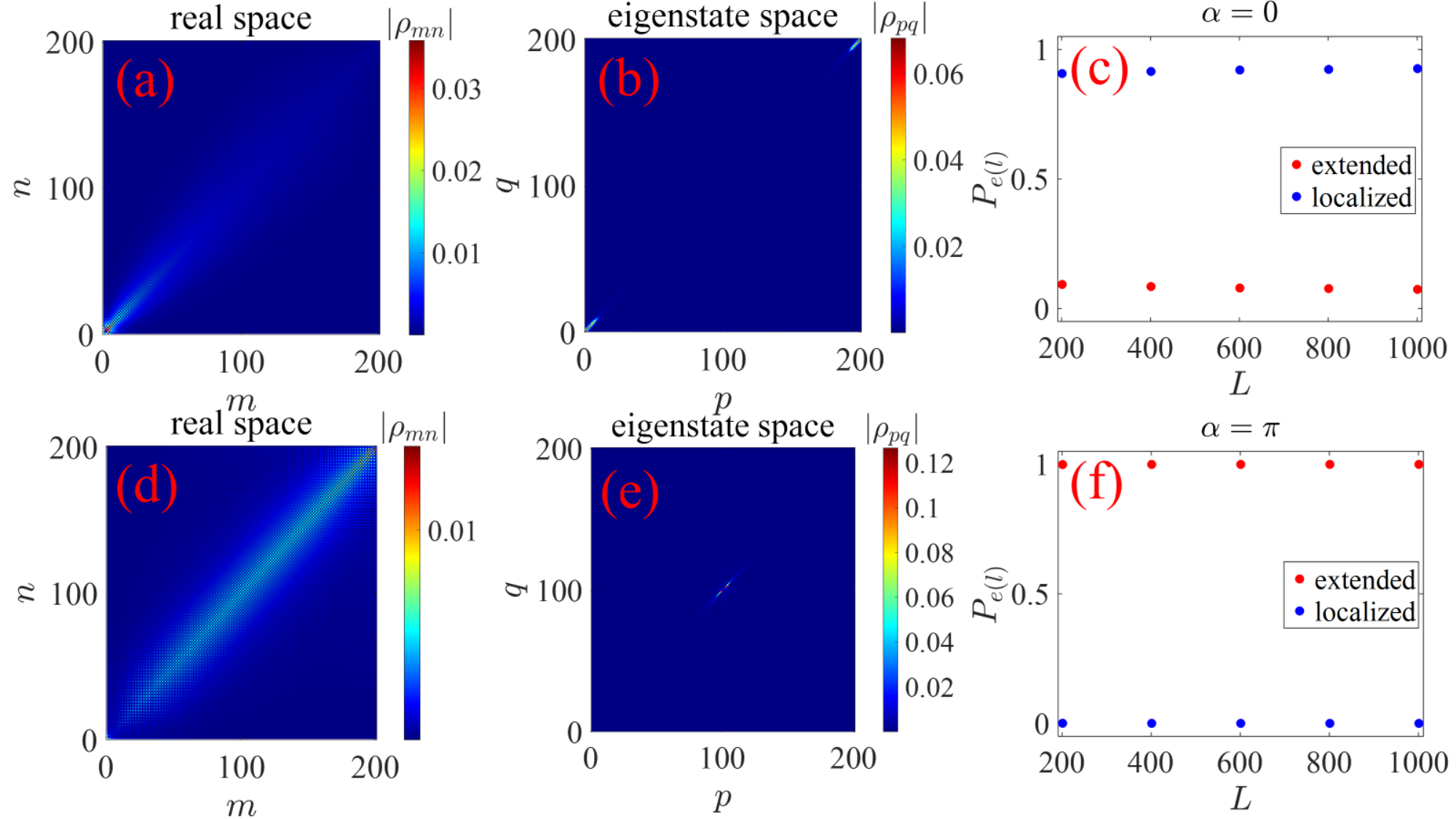}
        \caption{Localization properties of the stationary state for $\alpha=0$ and $\alpha=\pi$. (a),(d) Absolute values of the steady-state density-matrix elements $|\rho_{mn}|$ in real space for $\alpha=0$ and $\alpha=\pi$, respectively. (b),(e) Corresponding density-matrix distributions $|\rho_{pq}|$ in the eigenstate basis of the underlying Hamiltonian. (c),(f) Finite-size scaling of the eigenstate localization measure $P_e^{(l)}$ for system sizes $L=200$, $400$, $600$, $800$, and $1000$. The localized and extended states remain well separated over the entire size range for both values of $\alpha$, with no appreciable systematic drift with increasing $L$. The results demonstrate the robustness of the localized and extended sectors against the phase reversal of the dissipative channel. Other parameters are $l=2$, and $\Gamma=1$.}
	\label{fig3}
\end{figure*}

\section{Numerical results and discussion}\label{sec: results}

The spatial structure of the stationary state can be further characterized in both real space and the eigenstate basis of the underlying Hamiltonian. We first consider the case in which the jump operators in Eq.~\eqref{eq:jump_operator} have $l=1$ and analyze the steady state in the eigenstate representation of $H$. In this basis the density-matrix elements are $\rho_{mn}=\langle\psi_m|\rho_{\rm ss}|\psi_n\rangle$, where $|\psi_m\rangle$ and $|\psi_n\rangle$ denote eigenstates of $H$. Figures~\ref{fig2}(a) and \ref{fig2}(d) show the absolute values of the steady-state density-matrix elements $|\rho_{mn}|$ in real space for $\alpha=0$ and $\alpha=\pi$, respectively. In both cases the density matrix
exhibits a pronounced concentration along the diagonal $m=n$, accompanied by a spatially extended off-diagonal structure. The similarity between the two distributions indicates that the phase parameter $\alpha$ does not qualitatively modify the spatial structure of the stationary density matrix at the level of the real-space correlations resolved here. A complementary representation is obtained by transforming the stationary density matrix to the eigenbasis of $H$. As shown in Figs.~\ref{fig2}(b) and \ref{fig2}(e), the corresponding matrix elements
$|\rho_{pq}|$ remain concentrated within a well-defined region of eigenstate space for both $\alpha=0$ and $\alpha=\pi$. In particular, the dominant weight is centered near $p,q\simeq L/2$, indicating that the stationary state has substantial overlap with a localized eigenstate sector of the underlying disorder-free Hamiltonian. The pronounced concentration in eigenstate space thus provides a complementary characterization of the localization observed in real space.

To quantify the spatial character of the relevant eigenstates, we further examine the finite-size dependence of the eigenstate localization measure $P_{e,l}$ for different system sizes. By analyzing the diagonal elements of the density matrix, we can determine the proportions of extended and localized eigenstates in the steady-state for any given $\alpha$, denoted by $P_{e}=\sum_{n_{1}}\rho_{n_{1}n_{1}}$ ($P_{l}=\sum_{n_{2}}\rho_{n_{2}n_{2}}$), where $n_{1} (n_{2})$ denotes the index of extended (localized) eigenstates $|\psi_{n_{1}}\rangle$ ($|\psi_{n_{2}}\rangle$), respectively. For $\alpha=0$, the localized states, shown by the blue symbols in Fig.~\ref{fig2}(c), remain at $P_e\simeq0.85$--$0.9$ throughout the investigated size range, whereas the extended states, represented by the red symbols, remain close to $P_e\simeq0.1$. Importantly, neither branch exhibits a systematic drift with increasing $L$. The same behavior is observed for $\alpha=\pi$ [Fig.~\ref{fig2}(f)], where the localized and extended branches remain separated at approximately the same values over the entire range of system sizes.

The absence of a pronounced finite-size drift provides evidence that the distinction between the localized and extended sectors is not a finite-size artifact. In particular, the approximately size-independent behavior of $P_e^{(l)}$ for the localized branch is consistent with a robust localized character, while the well-separated extended branch remains characteristic of states that spread over the system. The persistence of this separation up to $L=1000$ indicates that the coexistence of localized and extended states is stable in the thermodynamic scaling regime accessible numerically. It is also noteworthy that the finite-size scaling is essentially unchanged upon switching the dissipative phase from $\alpha=0$ to $\alpha=\pi$. This observation is consistent with the phase-selective nature of the jump operator: changing $\alpha$ modifies the relative phase selected by the dissipative channel, while the underlying distinction between localized and extended sectors remains robust. Thus, the nonlocal dissipation provides control over the phase structure of the stationary state without eliminating the localization properties inherited from the disorder-free hopping-modulated lattice.

Taken together, Figs.~\ref{fig2}(a)--\ref{fig2}(f) establish the coexistence of localized and extended sectors from three complementary perspectives: the real-space structure of the stationary density matrix, its representation in the eigenstate basis, and the finite-size scaling of the eigenstate localization measure. These results demonstrate that the localization properties of the disorder-free lattice persist in the presence of the phase-selective dissipative dynamics. We emphasize that the best way to characterize the steady state's localization properties is in the eigenbasis of the Hamiltonian. The localization property of the steady state can be detected by measuring the system after dissipation is removed. Upon reaching a steady state, if dissipation is subsequently eliminated, the diagonal elements of the density matrix remain constant, as the dynamics are governed by the Hamiltonian via the equation $\frac{d \rho(t)}{dt} =  -i [H, \rho(t)]$. Consequently, the fraction of localized or delocalized eigenstates remains unchanged over time, reflecting the localization properties of the steady state.

\begin{figure}[t!]
	\centering
	\includegraphics[width=0.48\textwidth]{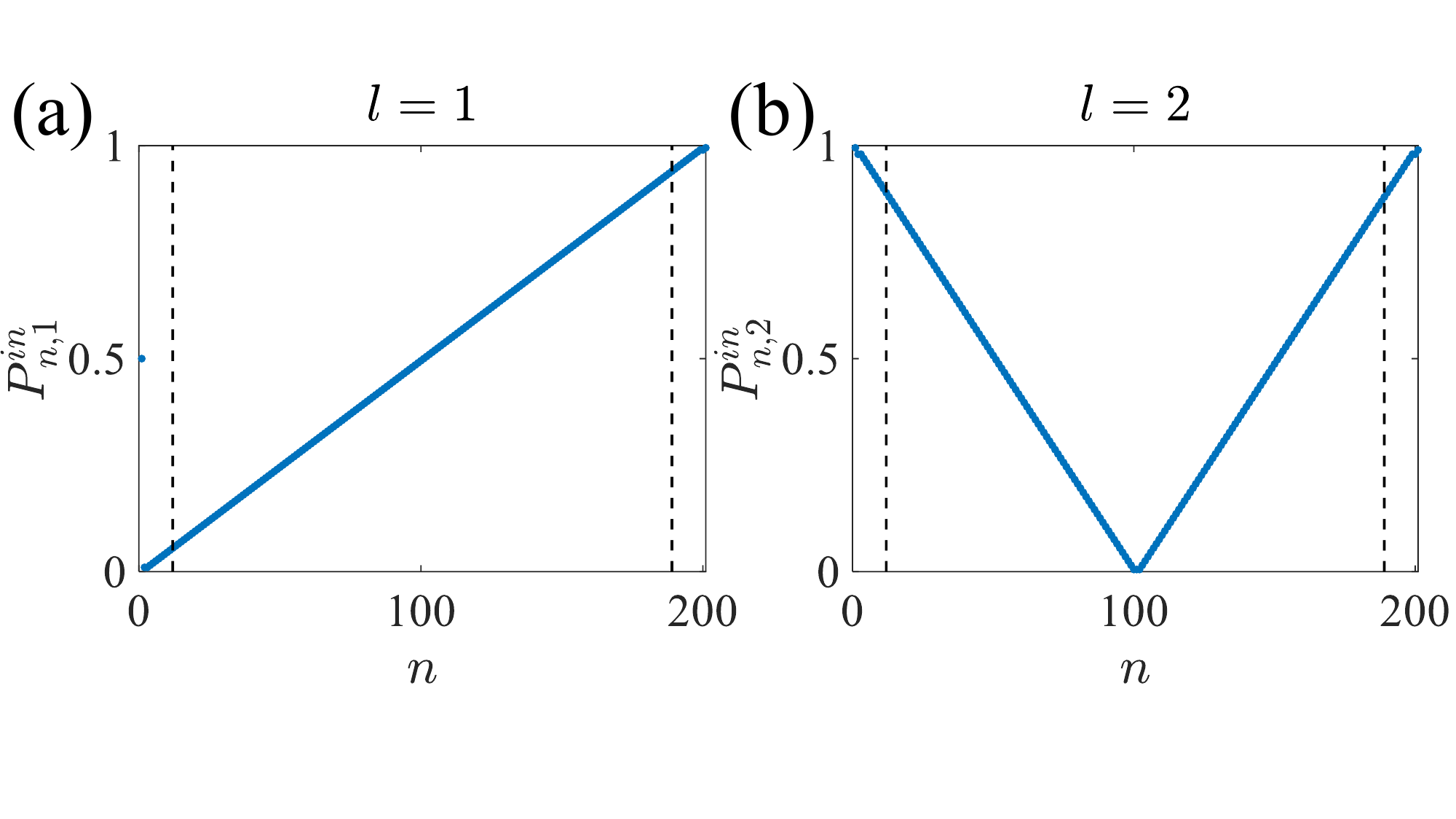}
        \vspace{-0.25in}
        \caption{(a) The fraction of in-phase pairs $P_{n,1}^{\mathrm{in}}$ for different eigenstates. The black dashed lines mark the extended bulk and localized boundary regions as eigenvalues increase. (b) The fraction of in-phase pairs $P_{n,2}^{\mathrm{in}}$ for different eigenstates.}
	\label{fig4}
\end{figure}

To elucidate the origin of steady states composed of extended or localized states, we examine the relative phases between pairs of sites separated by $l$. For the $n$-th eigenstate $|\psi_n\rangle=\sum_{j}\psi_{n,j}c_j^{\dagger}|0\rangle$, we define the phase difference between sites $j$ and $j+l$ as $\Delta\theta_{j,l}^{(n)}=\arg(\psi_{n,j})-\arg(\psi_{n,j+l})$. A pair is classified as in-phase when $\Delta\theta_{j,l}^{(n)}=0$. Counting the number of in-phase pairs $N_{n,l}^{\mathrm{in}}$, we define their fraction as $P_{n,l}^{\mathrm{in}}=\frac{N_{n,l}^{\mathrm{in}}}{N_{\mathrm{tot}}}, \qquad N_{\mathrm{tot}}=L-l$, where $N_{\mathrm{tot}}$ is the total number of site pairs separated by
$l$. For $l=1$, Fig.~\ref{fig4}(a) shows that $P_{n,1}^{\mathrm{in}}$ increases monotonically with the eigenstate index
$n$ (ordered by increasing energy), from approximately zero at the low-energy band edge to unity at the high-energy band edge. Thus, low-energy (high-energy) eigenstates carry small (large) in-phase fractions. This energy dependence accounts for the composition of the steady state: because the dissipative phase $\alpha$ selects eigenstates with a specific in-phase structure, the steady state for $\alpha=\pi$ ($\alpha=0$) is predominantly composed of localized boundary states from the low-energy (high-energy) region of the spectrum. We further consider jump operators with $l=2$ [Eq.~\eqref{eq:jump_operator}]. The corresponding in-phase fraction $P_{n,2}^{\mathrm{in}}$ exhibits a $V$-shaped profile
[Fig.~\ref{fig4}(b)]: it is maximal near both band edges and vanishes at the center of the spectrum. Accordingly, the localized states at both energy extrema contain predominantly in-phase site pairs, whereas the extended states in the middle of the spectrum carry predominantly out-of-phase pairs.

\begin{figure}[t!]
	\centering
	\includegraphics[width=0.48\textwidth]{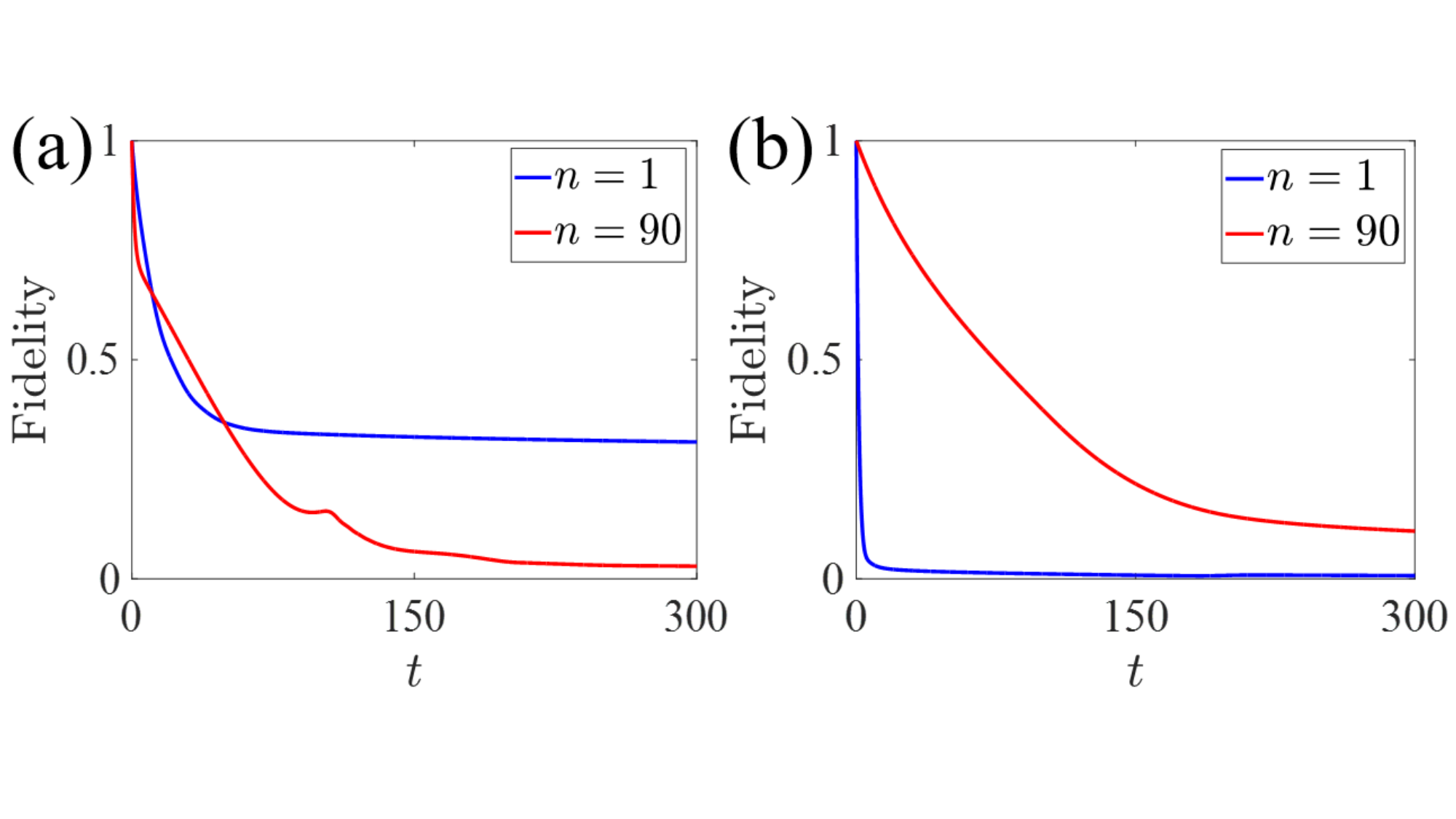}
        \vspace{-0.35in}
        \caption{Time evolution of the quantum fidelity $F[\rho(t),\rho_{0}]$ after the introduction of dissipation at $t=0$ with strength $\Gamma=1$. (a) and (d) correspond to the dissipative phases $\alpha=0$ and $\alpha=\pi$, respectively. In the figure, the blue solid line is obtained by initializing the system in the extended eigenstate labeled $n=1$, and the red solid line by initializing in the localized eigenstate labeled $n=90$. The system parameters are $L=201$ and $l=2$.}
	\label{fig5}
\end{figure}

We further examine this dissipation-tunable extended and localized steady states from a dynamical perspective. We prepare a localized or extended eigenstate as the initial state and introduce dissipation with $l=2$ at $t=0$. We then calculate the fidelity, which represents the overlap between the time-evolved state $\rho(t)$ and the initial state $\rho_{i}$, denoted as
\begin{equation}\label{eq:fidelity}
F\big[\rho(t),\rho_0\big] = \mathrm{Tr}\left[ \sqrt{\rho(t)^{1/2}\rho_0\rho(t)^{1/2}} \right].
\end{equation}
Starting with a few representative initial states, we compute the dissipative dynamics and evaluate the quantum fidelity. When the dissipation phase $\alpha=0$ [Fig.~\ref{fig5}(a)], the fidelity rapidly approaches zero for an extended initial state, indicating that the structure of the initial state has been completely modified. Conversely, the fidelity tends to a nonzero value for a localized initial state, suggesting that certain characteristics of the initial state are preserved. This is because the steady state is primarily composed of localized states, i.e., $P_{l}\gg P_{e}$. Finally, for $\alpha=\pi$ [Fig.~\ref{fig5}(b)], the steady state primarily consists of extended states. Therefore, the fidelity remains nonzero for an extended initial state but tends to zero for a localized one.

\section{Conclusion}\label{sec: conclusion}

In summary, we have shown that engineered bond dissipation provides deterministic, Hamiltonian-parameter-free control over the localization character of steady states in a strictly non-disordered one-dimensional lattice. The clean tight-binding model with deterministic inhomogeneous hopping supports a rich single-particle spectrum comprising extended bulk
states, exponentially localized boundary modes, and an algebraically localized bound state in the continuum. A number-conserving, nonlocal jump operator with a tunable relative phase $\alpha$ acts as a phase-selective decoherence channel: it favors eigenstates whose inter-site phase correlations match the dissipative bond pattern, without altering the
Hamiltonian spectrum.

The steady-state density matrix in the Hamiltonian eigenbasis reveals that varying $\alpha$ redistributes the asymptotic weight between localized and extended spectral sectors, for both $l=1$ and $l=2$ bond ranges. Finite-size scaling up to $L=1001$ confirms that the localized--extended separation is robust and not a finite-size artifact. The microscopic
mechanism underlying this selection is quantified by the in-phase pair fraction $P_{n,l}^{\mathrm{in}}$: for $l=1$ it varies monotonically with eigenenergy, whereas for $l=2$ it adopts a $V$-shaped profile peaked at both band edges. Dissipative quench dynamics, tracked through the quantum fidelity, further demonstrate that an initial extended or localized state is preserved or erased depending on whether its phase structure matches the selected sector. Crucially, once the steady state is reached and the dissipation is removed, the spectral weights remain frozen under unitary evolution, so the dissipatively prepared localization character is dynamically stable. Our results demonstrate that phase-selective bond dissipation is a general strategy for modulating steady-state states between local and extended regions. Extending this framework to higher-dimensional systems, interacting many-body mechanisms, and non-Markovian reservoirs provides a promising pathway for the preparation and transport engineering of dissipative states beyond the single-particle Markov limit.

\section*{Acknowledgments}

This work is supported by the National Natural Science Foundation of China (Grants No.~12304388, No.~12304290, and No.~12505017), the Beijing National Laboratory for Condensed Matter Physics (Grant No.~2025BNLCMPKF017), and the Fundamental Research Funds for the Central Universities.

\bibliography{Localization}
\end{document}